\documentclass[aps,pra,reprint,showpacs,superscriptaddress]{revtex4-1}
\usepackage{graphicx}
\usepackage{dcolumn}
\usepackage{bm}
\usepackage{amssymb}
\usepackage{amsmath}
\usepackage{subfigure}
\usepackage{epstopdf}
\usepackage{float}
\usepackage{mathrsfs}
\usepackage[colorlinks,citecolor=blue,
urlcolor=blue]{hyperref}
\usepackage{xcolor}
\begin{document}
\title{Skyrmion-based Magnetic Traps for Ultracold Atoms}
\author{Ren Qin}
\affiliation{School of Physics, Nankai University, Tianjin 300071, China}
\author{Yong Wang}
\email[]{yongwang@nankai.edu.cn}
\affiliation{School of Physics,	Nankai University, Tianjin 300071, China}

\begin{abstract}
We show that the stray field generated by isolated magnetic skyrmions can be used to trap ultracold atoms. Specially, ring-shaped and double-well trapping potentials for ultracold atoms can be created by combining the field from two isolated skyrmions. The geometry size, potential barrier, trapping frequency and Majorana loss rate of these magnetic traps can be tuned by the external magnetic field or device configuration. The results here could be useful to develop atomtronics devices by manipulating the magnetic skyrmions with modern spintronics techniques.
\end{abstract}

\maketitle

\section{Introduction}
The magnetic traps have been widely exploited to spatially confine and store the neutral atoms at extremely low temperatures, which are the key ingredients to investigate the ultracold atom physics and design atom-based quantum devices\cite{RMP2007,AtomChip}. The magnetic fields for the traps are usually generated from the current-carrying conductive microstructures\cite{RMP2007}, permanent magnets with fabricated patterns\cite{AtomChip,PRA2019}. Alternatively, several schemes have been proposed and realized to trap the ultracold atoms with the magnetic field carried by certain topological defects, such as the domain wall in ferromagnetic material\cite{APL2006,NL2012}, vortex in superconductors\cite{NJP2010,PRL2013}, \emph{etc}. Compared with the artificial structures, the magnetic traps based on topological defects can be controlled, reconfigured and scaled in an easier way.

The experimental discovery of magnetic skyrmion in chiral ferromagnetic materials\cite{Nature2006,Science2009,Nature2010} provides another promising opportunity to design the magnetic traps for ultracold atoms. The emergence of such kind of spin texture will significantly affect the stray field distribution near the film surface\cite{NC2018,APL2018,NJP2018}. Depending on the host materials, the size of magnetic skyrmion can vary from several nanometers to several hundreds nanometers\cite{NatNano2013,NatRevMat2016,NatRevMat2017}. This feature makes it flexible to get nanoscale magnetic traps, which is rather difficult for the conventional microstructures\cite{RMP2007,AtomChip}. Furthermore, the spintronics techniques to create, control, and eliminate the magnetic skyrmions have been well developed nowadays\cite{NatNano2013,NatRevMat2016,NatRevMat2017}. Therefore, the skyrmion-based magnetic traps can be easily reconfigured and moved by modern spintronics techniques, which will be advantageous for practical applications. If fact, it has been proposed that several types of magnetic lattices for ultracold atoms can be constructed from the stray field of chiral ferromagnetic film in skyrmion lattice phase\cite{PRA2019-2}.

In this paper, we show that the magnetic traps can be constructed from the stray field of isolated magnetic skyrmions. Furthermore, we find that the range-shaped and double-well magnetic traps can be realized by placing two magnetic skyrmions appropriately. We will also study how to tune the physical properties of these traps by external magnetic field or device geometry, and discuss their potential applications to develop quantum devices based on ultracold atoms. 

\begin{figure}[H]
\includegraphics[width=\linewidth]{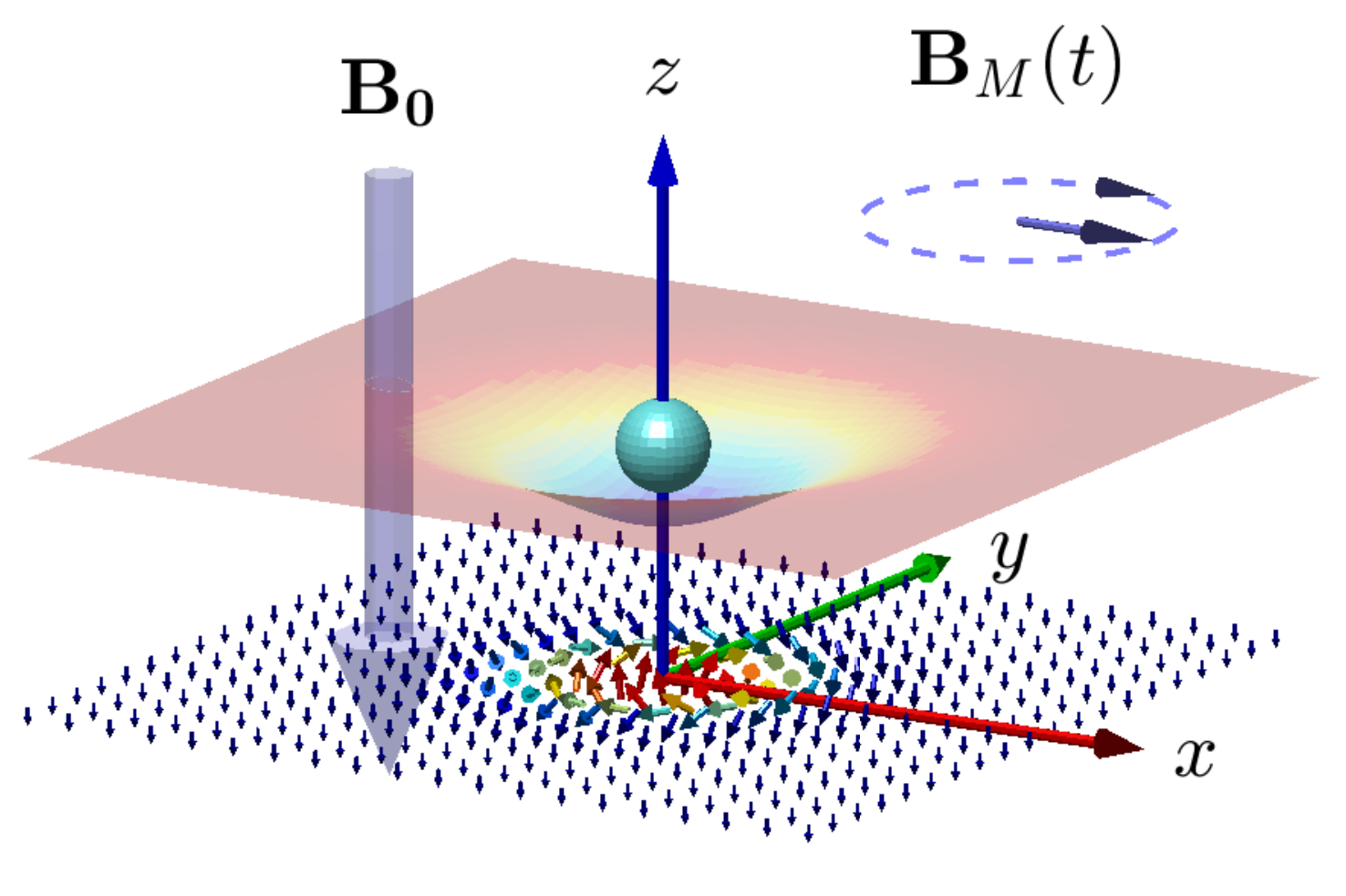}
\caption{(Color online) Schematic diagram for the magnetic trap with single skyrmion. A chiral ferromagnetic film is placed in the $x$-$y$ plane with an isolated skyrmion at the origin of coordinates. $\textbf{B}_{0}=(0,0,-B_{0})$ is a bias magnetic field to create a zero-field point at finite height, and $\textbf{B}_{M}(t)=B_{M}(\sin(\omega_{M}t),\cos(\omega_{M}t),0)$ is a rotating magnetic field to suppress the Majorana loss. An untracold atom is trapped at the minimum point of the magnetic potential.}\label{Fig1}
\end{figure}

\section{Trap with Single Skyrmion}
Fig.~\ref{Fig1} demonstrates the basic principle to construct a magnetic trap for ultracold atoms from one isolated magnetic skyrmion. The skyrmion is generated in a chiral ferromagnetic film with thickness $d$, which will establish the stray field $\mathbf{B}_{c}(\mathbf{r})$ near the film surface. Recently, this stray field has been measured by nitrogen vacancy center based magnetometry\cite{NC2018} and magnetic force microscopy\cite{APL2018}. A bias magnetic field $\mathbf{B}_{0}$ is applied in order to create a zero-field point at finite height. Such a zero-field point is desirable to trap the ultracold atoms prepared at weak-field seeking state\cite{AtomChip}. Besides, a rotating magnetic field $\mathbf{B}_{M}=B_{M}(\sin(\omega_{M}t),\cos(\omega_{M}t),0)$ is introduced to suppress the Majorana loss at the zero-field point\cite{PRL2013}. Such a configuration will generate a potential $U(\mathbf{r})=m_{F}g_{F}\mu_{B}B(\mathbf{r})$ for the atom in the hyperfine state $|F,m_{F}\rangle$. Here, $B(\mathbf{r})$ is the modulus of the total field $\mathbf{B}=\mathbf{B}_{c}+\mathbf{B}_{0}+\mathbf{B}_{M}$; $g_{F}$ is the Land\'{e} factor, $\mu_{B}$ is the Bohr magneton, $F$ is the total angular momentum quantum number, $m_F$ is the magnetic quantum number of the atom. This potential can trap the ultracold atoms with weak-field seeking state ($m_{F}g_{F}>0$) at its minimum point, with the trapping frequency $\omega_{T}=\sqrt{U''/m_{a}}$ and the Majorana loss rate $\Gamma/2\pi=\omega_{T}\exp(-4\omega_{L}/\omega_{T})$\cite{AtomChip,PRL2013}. Here, $m_{a}$ is the mass and $\omega_{L}$ is the Larmor precession frequency of the atom. 

The magnetization configuration $\mathbf{m}(\mathbf{r})$ for a single magnetic skyrmion can be obtained by performing the microscopic magnetic simulations for the chiral ferromagnetic film, whose energy density functional is defined as\cite{NJP2018,NatNano2013,NatRevMat2016,NatRevMat2017,PRA2019-2}
\begin{eqnarray}
\mathcal{E}[\mathbf{m}]=\frac{J}{2}(\nabla\mathbf{m})^{2}-D\mathbf{m}\cdot(\nabla\times\mathbf{m})-K\mathbf{m}_{z}^2-M_{s}\mathbf{B}_{0}\cdot\mathbf{m}.\nonumber\\\label{Em}
\end{eqnarray}
The four terms in $\mathcal{E}[\mathbf{m}]$ describes the Heisenberg exchange interaction, Dzyaloshinskii-Moriya interaction, perpendicular magnetic anisotropy energy, and Zeeman energy, respectively. During all the simulations in this paper, the size of the film is set as $100\times 100\times 1$~nm$^{3}$ and periodic boundary condition has been exploited. The material parameters are set as $J=20$~pJ/m, $D=3$~mJ/m$^2$, $K=0.7$~MJ/m$^3$. Besides, the saturation magnetization $M_{s}=M s = 580$~kA/m and the Gilbert damping coefficient $\alpha=0.05$ have been used. Then the profile of an isolated magnetic skyrmion is obtained by relaxing an initial configuration $\mathbf{m}_{0}(\mathbf{r})$ of the film, where a $21\times 21\times 1$~nm$^{3}$ region in the middle is set as (0,0,1) and the other part as (0,0,-1). Finally, the spatial distribution of the stray field $\mathbf{B}_{c}(\mathbf{r})$ is calculated from the magnetization configuration $\mathbf{m}(\mathbf{r})$ as
\begin{eqnarray}
\mathbf{B}_{c}(\mathbf{r})=\frac{\mu_0}{4\pi}\int d\mathbf{r'}\frac{3(\mathbf{m}(\mathbf{r}')\cdot\hat{\mathbf{R}})\hat{\mathbf{R}}-\mathbf{m}(\mathbf{r}')}{R^3}.\label{Bc}
\end{eqnarray}
Here, $\mu_{0}$ is the vacuum permeability, and $\mathbf{R}=\mathbf{r}-\mathbf{r}'$ denotes the displacement vector between two spatial points. 

\begin{figure}
\includegraphics[width=\linewidth]{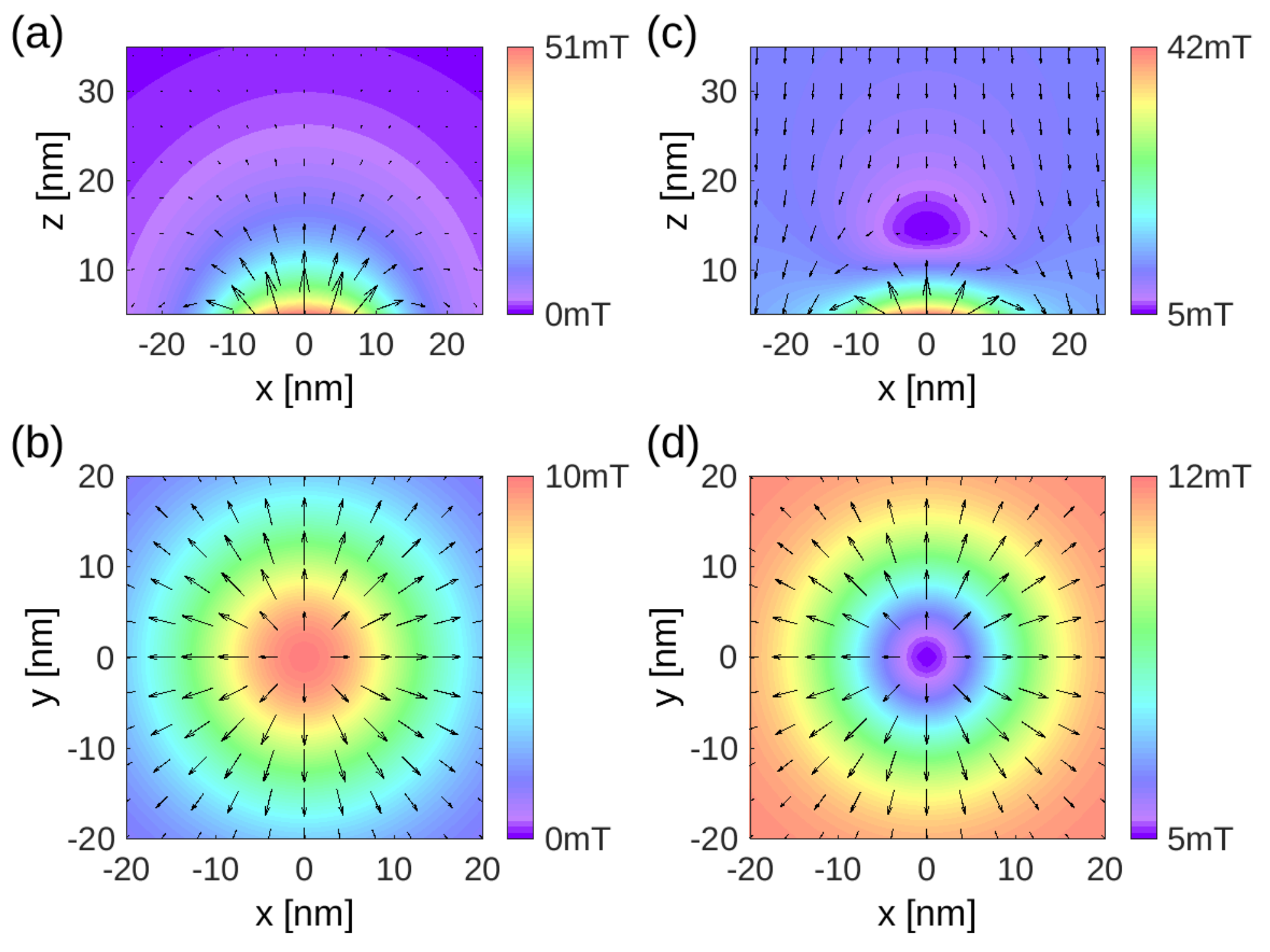}
\caption{(Color online) Magnetic field distributions of the magnetic trap with single skyrmion. (a) $\textbf{B}_{c}$ and its modulus in the $x$-$z$ plane with $y=0$~nm; (b) $\textbf{B}_{c}$ and its modulus in the $x$-$y$ plane with $z=15$~nm; (c) $\textbf{B}$ and its modulus in the $x$-$z$ plane with $y=0$~nm; (d) $\textbf{B}$ and its modulus in the $x$-$y$ plane with $z=15$~nm. Here, $\textbf{B}_{c}$ is numerically calculated from the microscopic simulation; $B_{0}=10$~mT and $B_{M} = 5$~mT have been exploited to construct the trap.}\label{Fig2}
\end{figure}

After setting the skyrmion center as the origin of coordinates and the chiral ferromagnetic film in the $x$-$y$ plane, Fig.~\ref{Fig2}(a) and (b) present the distribution of $\mathbf{B}_{c}(\mathbf{r})$ in the cross-sectional planes $y=0$~nm and $z=15$~nm when $\mathbf{B}_{0}=0$. As expected, $\mathbf{B}_{c}(\mathbf{r})$ is cylindrically symmetric around the $z$-axis and will gradually decay away from the skyrmion. Especially, $\mathbf{B}_{c}(\mathbf{r})$ on the $z$-axis will be along the $z$-direction. These features can be further understood from an approximate expression of $\mathbf{m}(\mathbf{r})=(\sin\theta\cos\psi,\sin\theta\sin\psi,\cos\theta)$ in the polar coordinates for the isolated skyrmion \cite{APL2018}, where 
\begin{eqnarray}
\theta&=&\pi-\sin^{-1}(\tanh(\frac{\rho+R}{w}))-\sin^{-1}(\tanh(\frac{\rho-R}{w})),\nonumber\\
\psi&=&\tan^{-1}(\frac{y}{x}).\nonumber
\end{eqnarray}
Here, $\rho=\sqrt{x^{2}+y^{2}}$, $R$ and $w$ are the radius and thickness of the skyrmion\cite{APL2018}. With the dipole approximation\cite{APL2018}, the stray field in the cylindrical coordinates generated by the skyrmion will be
\begin{eqnarray}
B_{c,\rho}=\frac{\mu_{0}P}{4\pi}\frac{3\rho z}{(\rho^{2}+z^{2})^{5/2}},\quad
B_{c,z}=\frac{\mu_{0}P}{4\pi}\frac{2z^{2}-\rho^{2}}{(\rho^{2}+z^{2})^{5/2}}.\nonumber
\end{eqnarray}
Here, $\mathbf{P}=P\hat{\mathbf{z}}$ is the magnetic dipole moment of the skyrmion with $P=2\pi M_{s}d\int_{0}^{\infty}d\rho\rho(1+\cos\theta(\rho))$\cite{APL2018}.
When $\rho=0$, we will get $B_{c,\rho}=0$ and $B_{c,z}=\frac{\mu_{0}}{2\pi}Pz^{-3}$, which coincides with the numerical results in Fig.~\ref{Fig1}(a) and (b) above.

When the bias magnetic field $\mathbf{B}_{0}=(0,0,-B_{0})$ is turned on, the zero-field point will be created on the $z$-axis with the height $z_{\text{min}}=(\frac{\mu_{0}}{2\pi}\frac{P}{B_{0}})^{\frac{1}{3}}$. The application of rotating magnetic field $\mathbf{B}_{M}$ will further turn the zero-field point into non-zero minimum point in the trapping potential $U(r)$. Fig.~\ref{Fig2}(c) and (d) show the field distribution in the cross-sectional planes $y=0$~nm and $z=15$~nm after setting $B_{0}=10$~mT and $B_{M}=5$~mT. Thereby, a magnetic trap has been established to trap the ultracold atoms in weak-field seeking state.

\begin{figure}
\includegraphics[width=0.9\linewidth]{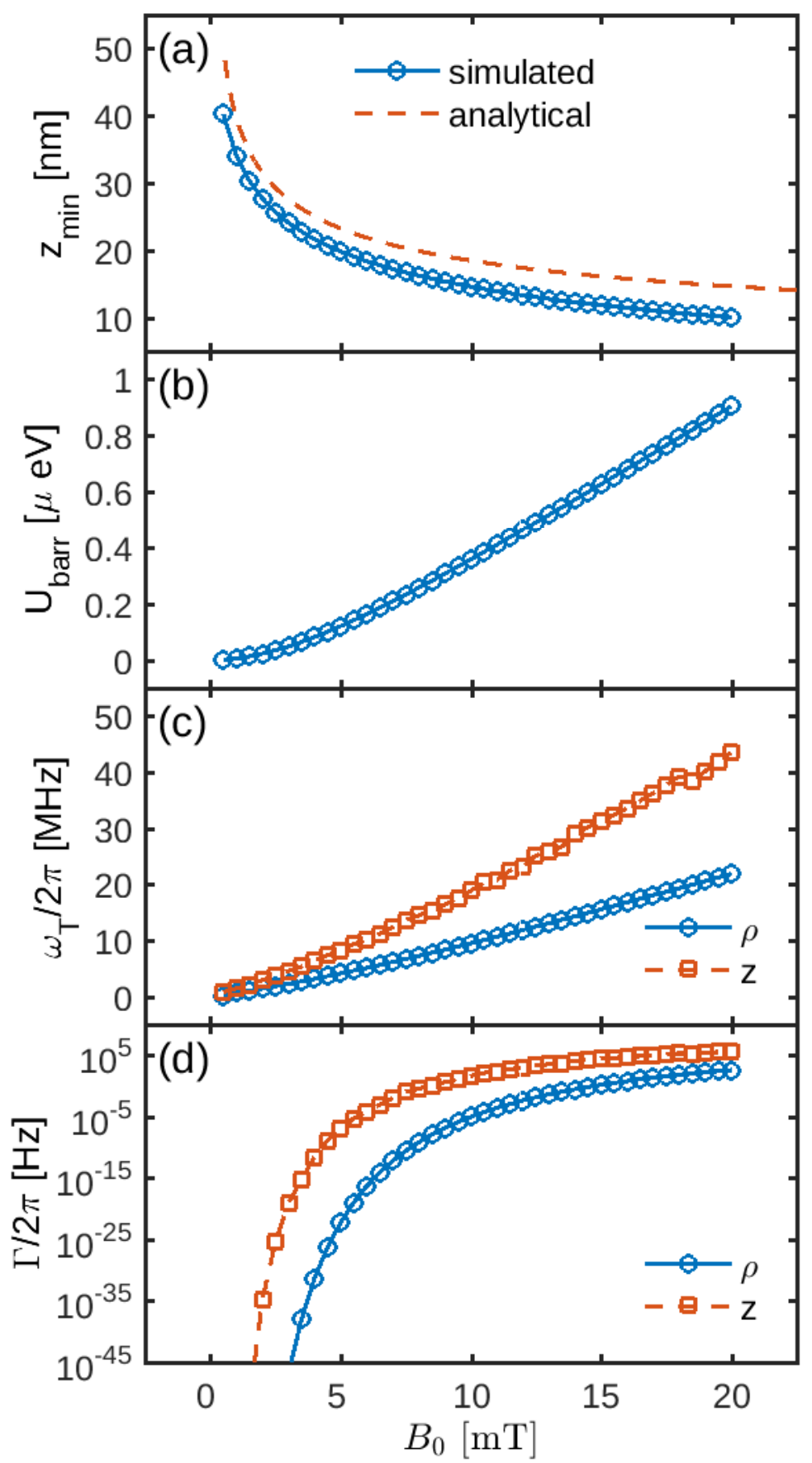}
\caption{(Color online) The dependence of trap parameters on the bias field $B_{0}$ for $^{87}$Rb atom in the hyperfine state $|F=2,m_{F}=2\rangle$. (a) trapping height $z_{min}$, which has been obtained from micromagnetic simulation (real line) and from the dipole approximation analytically (dashed line); (b) trapping barrier $U_{barr}$; (c) trapping frequency $\omega_{T}$; (d) Majorana loss rate $\Gamma$. Here, $B_{M}=5$~mT. $\rho$ and $z$ denote the radical and axial components of $\omega_{T}$ and $\Gamma$.}\label{Fig3}
\end{figure}

The physical properties of the magnetic trap can be further tuned by the bias magnetic field $\mathbf{B}_{0}$. Firstly, the trapping height $z_{min}$ will become smaller with stronger bias field. As shown in Fig.~\ref{Fig3}(a), $z_{min}$ extracted from the simulated $U(\mathbf{r})$ will decline from about 40 to 10~nm when $B_{0}$ increases from $0.5$ to $20$~mT, which also agree well with the analytical results in the dipole approximation. This fact makes it possible to load the cold atoms from infinite region to the magnetic trap adiabatically\cite{PRL1998}.   

Besides, the trapping barrier $U_{barr}$, trapping frequency $\omega_{T}$ and Majorana loss rate $\Gamma$ of the trap are investigated for $^{87}$Rb atom in the hyperfine state $|F=2,m_{F}=2\rangle$. The trapping barrier $U_{barr}=U(r\rightarrow\infty)-U(r_{min})$ is the potential for the atoms to overcome. Considering that $B(r\rightarrow\infty)=\sqrt{B_{0}^{2}+B_{M}^{2}}$ and $B(r_{min})=B_{M}$, $U_{barr}$ will be irrelevant to the magnetic field $\textbf{B}_{c}$, which  vanishes far away from the magnetic skyrmion. As shown in Fig.~\ref{Fig3}(b), $U_{barr}$ will monotonically increase from $0$ to about $1$~$\mu$eV along with the increased bias field $B_{0}$ from $0$ to $20$~mT. Since the trapping potential is cylindrically symmetric around the $z$-axis, there will exist the radical and axial components for both $\omega_{T}$ and $\Gamma$. As shown in Fig.~\ref{Fig3}(c) and (d), when the bias field $B_{0}$ increases from $0$ to 20~mT, $\omega_{T,\rho}/2\pi$ will increase from $0$ to about 40~MHz and $\omega_{T,z}/2\pi$ will increase from $0$ to about $20$~MHz; meanwhile, $\Gamma/2\pi$ will exponentially increase to about $10^{5}$~Hz due to the large trapping frequency. Therefore, a stronger bias field can result in a higher trapping frequency for the trap, but will also cause larger Majorana loss. A moderate bias magnetic field should be chosen in practical applications.

\section{Ring-shaped and Double-well magnetic Traps}

\begin{figure}
\includegraphics[width=\linewidth]{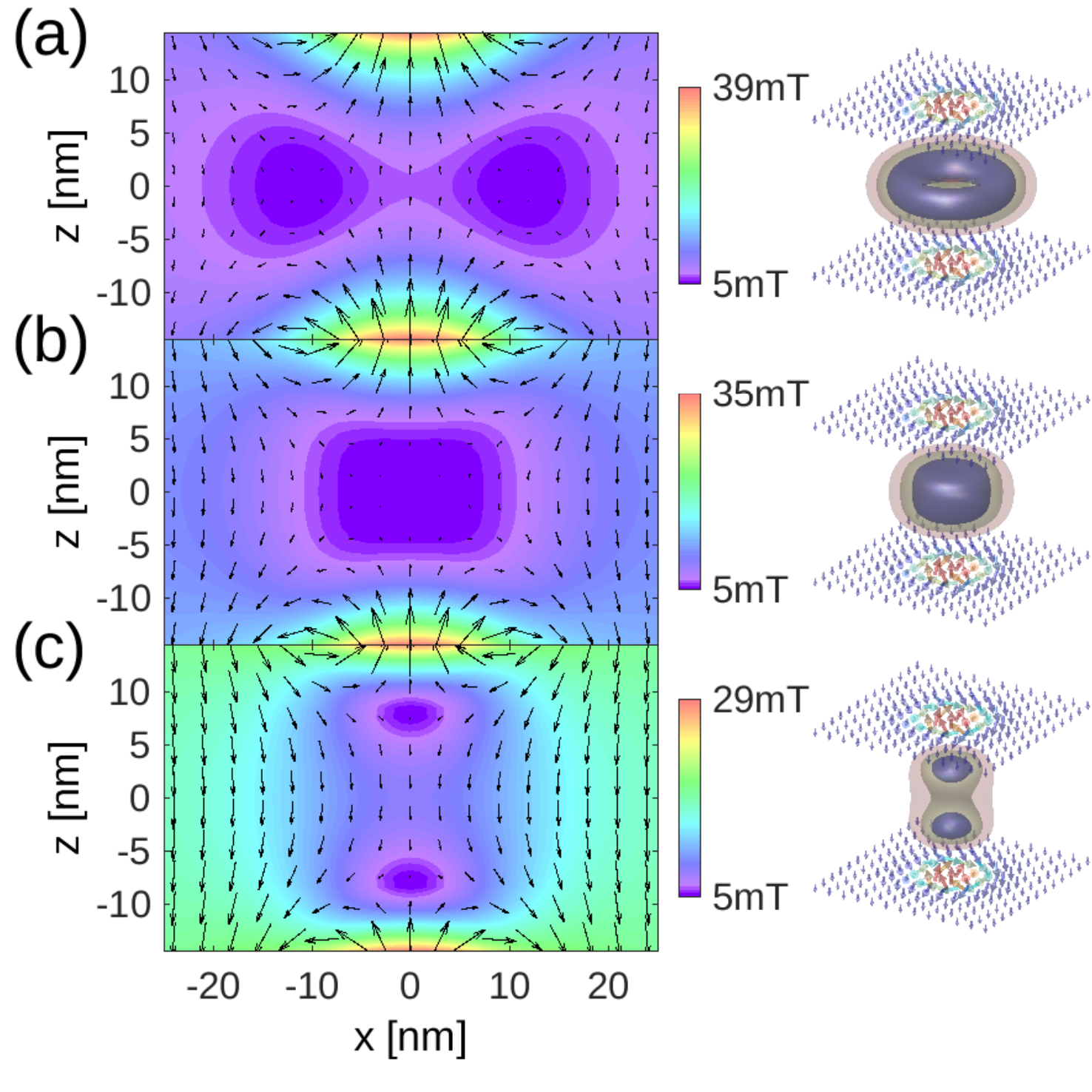}
\caption{(Color online) Magnetic field distributions in the $x$-$z$ plane with $y=0$~nm (left) and isosurfaces of field modulus $B(\textbf{r})$ (right) for three types of magnetic traps formed by two isolated skyrmions. (a) ring-shaped trap with isosurface values $5.5, 6, 6.5$~mT; (b) intermediate trap with isosurface values $6, 7, 8$~mT; (c) double-well trap with isosurface values $7, 9, 11$~mT. The bias fields $B_{0}$ for the three traps are $5$,$9$,$15$~mT, respectively. The distance $d$ between the two skyrmions is $41$~nm, and $B_{M}=5$~mT.}\label{Fig4}
\end{figure}

In addition to the single-well trap, other types of traps have also been designed and realized for ultracold atoms. Examples include the ring-shaped trap\cite{PRL2001,PRL2005,PRA2006,NJP2008} for inertial sensor or gyroscope and double-well trap\cite{PRA2001,PRL2004,NP2005} for matter-wave interferometry. Here, we show how to construct ring-shaped and double-well magnetic traps with two magnetic skyrmions locating in two identical chiral ferromagnetic films. These two films are in parallel to the $x$-$y$ plane and their distance is $d$. The centers of the two skyrmions are both on the $z$-axis, and the origin of the coordinates is reset in the middle of the two centers. Then the cylindrical symmetry around the $z$-axis will still be maintained for the stray field $\mathbf{B}_{c}$ between the two films. Within the dipole approximation, the field distribution in the $z=0$ plane will be
\begin{eqnarray}
B_{c,\rho}(\rho,z=0)=0,\quad B_{c,z}(\rho,z=0)=\frac{\mu_{0}P}{2\pi}\frac{(2(\frac{d}{2})^{2}-\rho^{2})}{(\rho^{2}+(\frac{d}{2})^{2})^{\frac{5}{2}}};\nonumber
\end{eqnarray}
while in the $z$-axis, one has
\begin{eqnarray}
B_{c,\rho}(\rho=0,z)=0,\quad B_{c,z}(\rho=0,z)=\frac{\mu_{0}P}{\pi}\frac{({(\frac{d}{2})}^3+3(\frac{d}{2})z^2)}{{({(\frac{d}{2})}^2-z^2)}^3}.\nonumber
\end{eqnarray}
Obviously, the zero-field points of $\mathbf{B}_{c}$ will form a ring with the radius $\rho_{min}=\frac{d}{\sqrt{2}}$ in the $z=0$ plane if no bias magnetic field is applied. Therefore, the ultracold atoms in the weak-field seeking state can be trapped in a ring-shaped potential in this case. However, if a bias magnetic field $\mathbf{B}_0=(0,0,-B_{0})$ is present, the radius $\rho_{min}$ of the ring will expand if $B_{0}<0$ and shrink if $B_{0}>0$. In particular, the ring will shrink into a point at the critical value $B_{0}=\frac{8\mu_{0}P}{\pi d^{3}}$, where the ring-shaped trap will become a single magnetic trap. Furthermore, if $B_{0}>\frac{\mu_{0}}{\pi}\frac{8P}{d^{3}}$, two zero-field points will symmetrically appear at $\pm z_{min}$ on the $z$-axis, which can be used to construct a double-well magnetic trap for cold atoms.

\begin{figure}
\includegraphics[width=\linewidth]{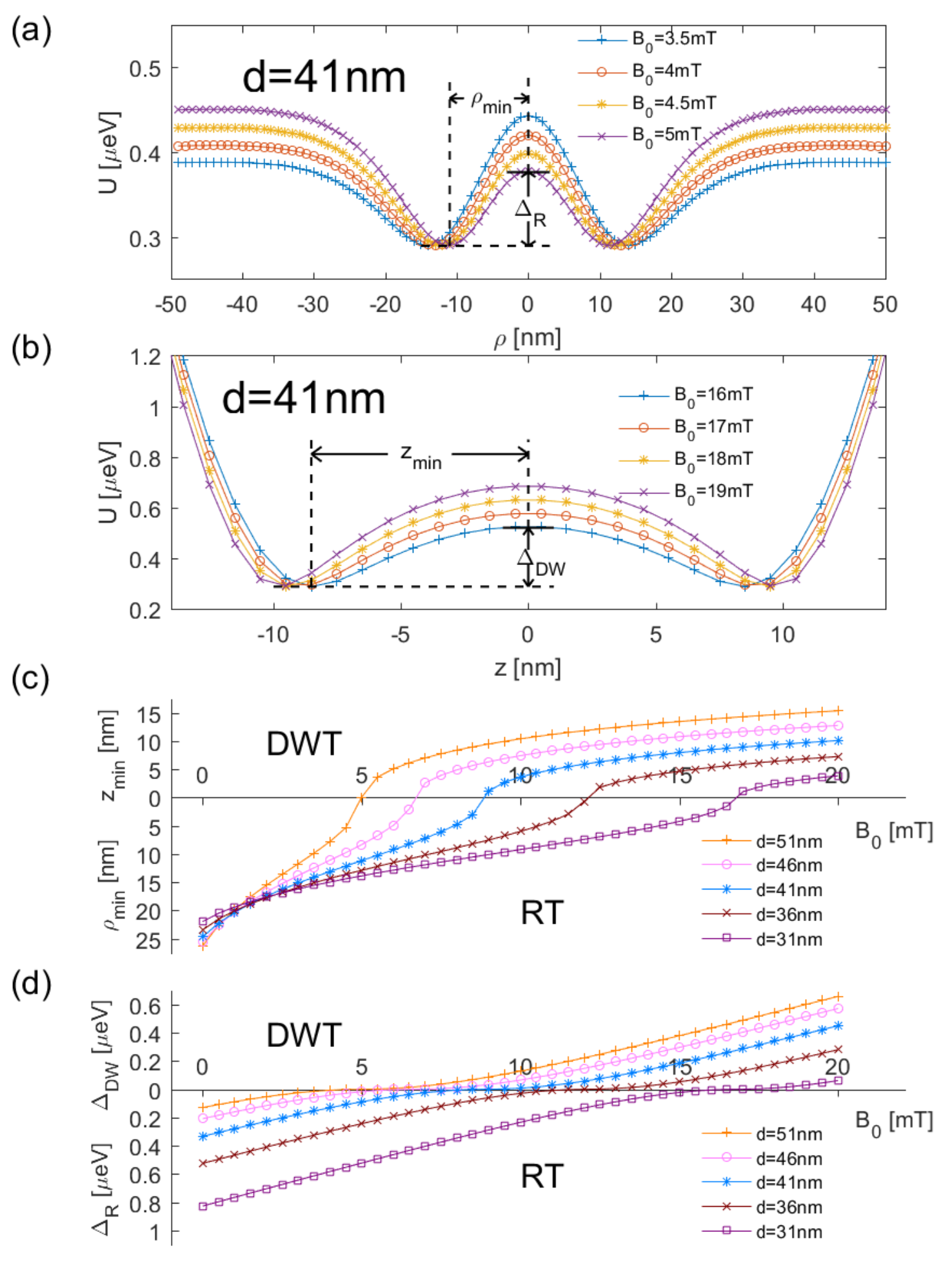}
\caption{(Color online) The dependence of trapping potential $U(\textbf{r})$ on the bias field $B_{0}$ and distance $d$ for $^{87}$Rb atom in the hyperfine state $|F=2,m_{F}=2\rangle$. (a) $U(\rho,z=0)$ of the ring-shaped trap with fixed $d=41$~nm and different $B_{0}=3.5,4,4.5,5$~mT; (b) $U(\rho=0,z)$ of the double-well trap with fixed $d=41$~nm and different $B_{0}=16,17,18,19$~mT; (c) $B_{0}$-dependent trap size $\rho_{min}$ and $z_{min}$ for different $d=31,36,41,46,51$~nm; (d) $B_{0}$-dependent potential barrier $\Delta_{R}$ and $\Delta_{DW}$ for different $d=31,36,41,46,51$~nm. RT: ring-shaped trap; DWT: double-well trap.  }\label{Fig5}
\end{figure}

The deduction above is confirmed by the spatial distribution of magnetic field between two chiral ferromagnetic films based on micromagnetic simulation. The same material parameters in previous section have been exploited and the distance of the two films is $d=41$~nm. A rotating magnetic field with $B_{M}=5$~mT is also included. As an example, Fig.~\ref{Fig4}(a) displays the field distribution in the cross-sectional plane $y=0$~nm for a ring-shaped trap, where the bias field is $B_{0}=5$~mT. When $B_{0}$ is increased to $9$~mT, the ring-shaped trap will becomes a single magnetic trap centered at the origin with its field distribution in Fig.~\ref{Fig4}(b). This single magnetic trap will then be split into the double-well trap for larger bias field, and the case for $B_{0}=15$~mT is given in Fig.~\ref{Fig4}(c). 

The evolution of magnetic potential $U(\mathbf{r})$ under the bias field $B_{0}$ is further calculated for $^{87}$Rb atom in the hyperfine state $|F=2,m_{F}=2\rangle$. For given $d=41$~nm, Fig.~\ref{Fig5}(a) shows $U(\rho,z=0)$ for the ring-shaped traps with $B_{0}=3.5,4.0,4.5,5.0$~mT, while Fig.~\ref{Fig5}(a) shows $U(z,\rho=0)$ for the double-well traps with $B_{0}=16,17,18,19$~mT. The results show that the order of magnitude of $U(\mathbf{r})$ will be tenths of $\mu$eV, which corresponds the temperature at the level of mK. Moreover, the ring-shaped trap and double-well trap are characterized by the size $\rho_{min},z_{min}$ and the potential barrier $\Delta_{R}$,$\Delta_{DW}$, and their dependence on the bias field $B_{0}$ and the distance $d$ are given in Fig.~\ref{Fig5}(c) and (d), respectively. For fixed $d$, $r_{min}$ will continuously reduce to zero and then $z_{min}$ will monotonically increase from zero when $B_{0}$ is continuously increased from 0 to 20~mT, which coincide with the analytical results above. Besides, the critical values of bias field where $r_{min}=z_{min}=0$ will become larger for smaller distance $d$. Therefore, it is also possible to switch the range-shaped and double-well magnetic traps by tuning the distance $d$ for fixed $B_{0}$. On the other hand, $\Delta_{R}$ will continuously decrease when the ring-shaped trap is shrunk, while $\Delta_{DW}$ will monotonically increase with the increased bias field. This fact can be further understood from the strength of total magnetic field at the origin, where $B(\mathbf{r}=0)=\sqrt{(\frac{8\mu_{0}P}{\pi d^{3}}-B_{0})^{2}+B_{M}^{2}}$ under the dipole approximation. At the critical field $B_{0}=\frac{8\mu_{0}P}{\pi d^{3}}$, one has $B(\mathbf{r}=0)=B_{M}$ and the potential barrier will vanish.

\begin{figure}
\includegraphics[width=\linewidth]{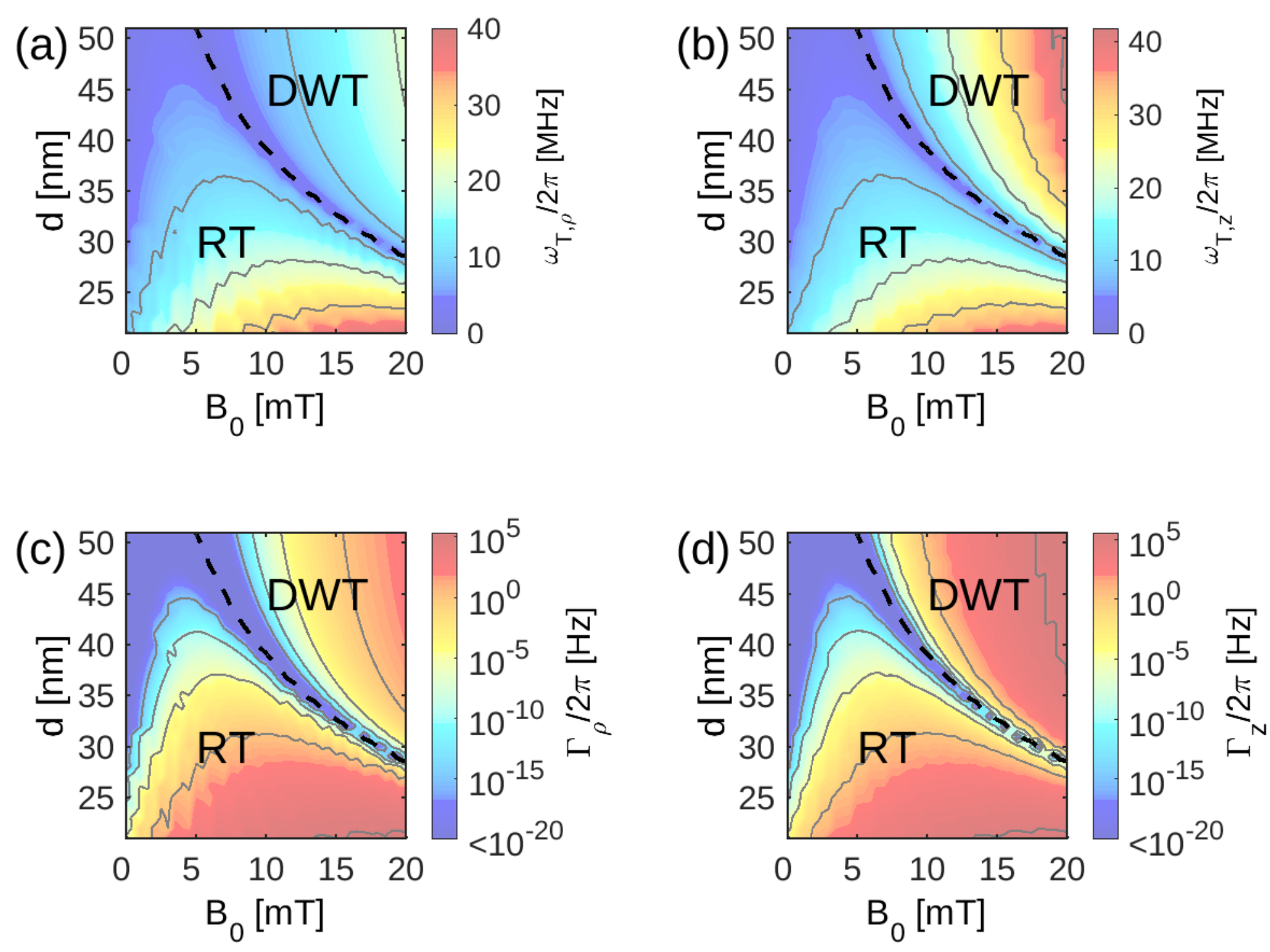}
\caption{(Color online) The dependence of trapping frequency $\omega_{T}$ and Majorana loss rate $\Gamma$ on the bias field $B_{0}$ and distance $d$ for $^{87}$Rb atom in the hyperfine state $|F=2,m_{F}=2\rangle$. (a) radical component of $\omega_{T}$; (b) axial component of $\omega_{T}$; (c) radical component of $\Gamma$; (d) axial component of $\Gamma$. RT: ring-shaped trap; DWT: double-well trap. The black dashed curve denotes the parameters to form the intermediate traps.}\label{Fig6}
\end{figure}

Fig.~\ref{Fig6} displays the trapping frequency $\omega_{T}$ and Majorana loss rate $\Gamma$ of the magnetic traps for $^{87}$Rb atom in the $B_{0}$-$d$ parameter plane, which is divided into two regions for the ring-shape trap and double-well trap by a critical curve for the single trap. As given in Fig.~\ref{Fig6}(a)(b), high $\omega_{T}$ can be achieved by increasing the bias field $B_{0}$ and decreasing the layer distance $d$ for the ring-shaped trap; while for the double-well trap, both large $B_{0}$ and $d$ will result in high $\omega_{T}$. On the other hand, $\omega_{T}$ will drastically decrease in the region around the critical curve or the line $B_{0}=0$ due to the loose trapping potential. Meanwhile, the Majorana loss rate will follow the same tread as the trapping frequency. As shown in Fig.~\ref{Fig6}(c)(d), $\Gamma/2\pi$ can vary from less than $0$ to $10^{5}$~Hz when $\omega_{T}/2\pi$ increases from $0$ to about $40$~MHz. The results here suggest that relatively high trapping frequency ($\sim 1$~MHz) with sufficiently low Majorana loss rate($\sim 1$~Hz) can be realized for both the ring-shaped and double-well magnetic traps.

\section{Conclusion}
In conclusion, we show that the stray field of isolated magnetic skyrmions in chiral ferromagnetic film can be utilized to design magnetic traps for ultracold atoms. Single-well trap, ring-shaped and double-well traps with high trapping frequency and low Majorana loss rate can be realized by reasonably choosing the material parameters, device geometry and external magnetic field. Considering that magnetic skyrmion is topologically stable and can be flexibly manipulated by modern spintronics techniques, the strategy proposed here could be advantageous to integrate the skyrmion-based magnetic traps into the atomtronics devices.  

\begin{acknowledgements}
This work has been supported by NSFC Projects No. 61674083 and No. 11604162 and by the Fundamental Research Funds for the Central Universities, Nankai University (Grant No. 7540).
\end{acknowledgements}

\end{document}